\documentclass[twocolumn,showpacs,prl]{revtex4}

\usepackage[dvips]{graphicx}

\renewcommand{\epsilon}{\varepsilon}
\newcommand{\figurewidth}{0.45\textwidth}

\begin{document}
\title{Stabilization of colloidal suspensions by means of highly-charged
nanoparticles}
\author{Jiwen Liu}
\author{Erik Luijten}
\email[Corresponding author. E-mail: ]{luijten@uiuc.edu}
\affiliation{Department of Materials Science and Engineering and Frederick
Seitz Materials Research Laboratory, University of Illinois at
Urbana-Champaign, 1304 West Green Street, Urbana, Illinois 61801, USA}

\date{September 22, 2004}

\begin{abstract}
  We employ a novel Monte Carlo simulation scheme to elucidate the
  stabilization of neutral colloidal microspheres by means of highly-charged
  nanoparticles [V. Tohver \emph{et al.}, Proc.\ Natl.\ Acad.\ Sci.\ U.S.A.
  \textbf{98}, 8950 (2001)]. In accordance with the experimental
  observations, we find that small nanoparticle concentrations induce an
  effective repulsion that prevents gelation caused by the intrinsic van der
  Waals attraction between colloids. Higher nanoparticle concentrations
  induce an attractive potential which is, however, qualitatively different
  from the regular depletion attraction. We also show how
  colloid--nanoparticle size asymmetry and nanoparticle charge can be used
  to manipulate the effective interactions.
\end{abstract}


\maketitle

Colloidal particles find widespread application as precursors for
nanostructured materials, including advanced coatings, drug carriers, and
colloidal crystals.  These materials are frequently fabricated from the
liquid phase via hierarchical self-assembly processes that are governed by
the interactions of the particles, their shape, and their size.  The
understanding and prediction of phase behavior and stability of a suspension
relies on a fundamental knowledge of the \emph{effective} forces between
colloids~\cite{russell89}, which arise from a combination of direct
interactions and indirect interactions mediated by the solvent and by other
solute particles.  Traditional methods to stabilize a suspension typically
involve tuning of the effective interactions through charged groups or by
grafting short polymer chains onto the colloidal surface.  These mechanisms,
however, pose serious problems in certain situations, such as the
fabrication of close-packed colloidal crystals, where they cause an increase
in the lattice spacing of the sedimented colloids and thus lead to cracking
of the crystal upon drying.

It is therefore of great fundamental and practical interest that an
alternative stabilization strategy has emerged~\cite{tohver01,tohver01b}.
In these experiments, the addition of very small concentrations of charged
zirconia nanoparticles was found to prevent the aggregation of near-neutral
silica microspheres in aqueous suspension. Upon further increase of the
zirconia concentration, microsphere aggregation was recovered, leading to a
``window of stability.''  Lewis and co-workers~\cite{tohver01} ascribe the
initial colloidal stabilization to the formation of a non-adsorbing ``halo''
of zirconia particles around the silica colloids, arising from the strong
electrostatic repulsion between nanoparticles.  However, while consistent
with zeta-potential measurements that confirm a weak accumulation of
nanoparticles near the colloidal surface, such halo formation is at variance
with the observation of stabilization at zirconia volume fractions below
$10^{-3}$, where the average nanoparticle separation is approximately
$40$nm, significantly larger than the electrostatic screening length
$\kappa^{-1} \approx 2$nm~\cite{tohver01b}.
On the other hand, the reentrant gelation is attributed to the regular
depletion attraction induced by the nanoparticles~\cite{asakura54}.  This is a
potentially contentious argument as well, as it has been demonstrated that the
depletion effect can be substantially modified by non-additivity
effects~\cite{mendez00,roth01,louis02} which, in this case, would result from
the electrostatic repulsions.

Unfortunately, theoretical and computational approaches experience severe
difficulties in the treatment of this system. Analytical methods commonly
incorporate fluctuation and correlation effects only in an approximate
manner. While simulations can explicitly account for these effects, inherent
limitations in most computational methodologies greatly restrict the range
of accessible size ratios in multi-component mixtures.  In this Letter, we
overcome this problem by exploiting a novel, highly-efficient Monte Carlo
scheme~\cite{geomc}, which permits the explicit inclusion of interacting
species of vastly different sizes.  Thus, we are able to determine the
distribution of nanoparticles and the colloidal potential of mean force as a
function of nanoparticle concentration.  Our findings constitute the first
\emph{quantitative} confirmation of the experimental observations and
clarify the underlying physical mechanism of the stabilization. In addition,
we provide explicit predictions for the role of two crucial parameters, the
magnitude of the nanoparticle charge and the colloid--nanoparticle size
asymmetry.

It is the purpose of our calculation to determine the effective pair potential
between colloidal particles induced by the nanoparticles.  This effective
interaction is independent of the intrinsic van der Waals attraction. If
sufficiently repulsive it can counteract the latter and thus stabilize the
suspension.  Both species are modeled as hard spheres with diameter
$\sigma_{\rm micro}$ and $\sigma_{\rm nano}$, respectively.  The aqueous
environment (water and screening ions) is represented as a homogeneous medium.
In addition, we take into account the electrostatic double-layer interactions
by means of the well-known Hogg--Healy--Fuerstenau (HHF)
formula~\cite{hogg66,sader95}.  For two nanoparticles at a surface
separation~$D$, this equation, under constant-potential conditions, reduces
to~\cite{hunter01}
\begin{eqnarray}
  V_{\rm nano-nano}
  &=&  \epsilon_0\epsilon_r\pi \sigma_{\rm nano}\Psi_{\rm nano}^2
  \ln[1+{\exp(-\kappa D)}] \nonumber\\
  &\approx&  \epsilon_0\epsilon_r\pi \sigma_{\rm nano}\Psi_{\rm nano}^2
  {\exp(-\kappa D)} \;,
  \label{eq:ele1}
\end{eqnarray} 
where the approximation is valid for $\kappa D \gg 1$.  $\Psi_{\rm nano}$ is
the zirconia zeta potential, $\epsilon_0$ the vacuum permittivity, and
$\epsilon_r = 80$ the dielectric constant of water.  Since the silica
spheres are negligibly charged, we ignore their electrostatic interaction.
However, an electrostatic double layer interaction arises between the
microspheres and the nanoparticles, which, for large size asymmetry $\alpha
\equiv \sigma_{\rm micro}/\sigma_{\rm nano} \gg 1$ is described
by~\cite{hogg66}
\begin{eqnarray}
  V_{\rm micro-nano}
  &=&  \frac{1}{2}\epsilon_0\epsilon_r\pi
  \sigma_{\rm nano}\Psi_{\rm nano}^2 \ln[1-{\exp(-2\kappa D)}]
  \nonumber\\
  &\approx&  
  - \frac{1}{2} \epsilon_0\epsilon_r\pi \sigma_{\rm nano}\Psi_{\rm nano}^2
  {\exp(-2\kappa D)} \;,
  \label{eq:ele2}
\end{eqnarray} 
where the approximation is again valid for $\kappa D \gg 1$.  While
charge-regulating boundary conditions are even more appropriate than
constant-potential boundary conditions for the interactions (\ref{eq:ele1})
and~(\ref{eq:ele2}), both conditions have been found to lead to comparable
results~\cite{behrens99}. In particular, the former can also yield an
attraction between neutral and charged surfaces~\cite{biesheuvel04}.
Indeed, whereas the presence of a microsphere--nanoparticle attraction is
not considered explicitly in Ref.~\cite{tohver01}, both the supernatant
measurements and the dynamic nature of the zeta potential measurement (which
implies a cooperative movement of colloids and nanoparticles) indicate a
certain degree of microsphere--nanoparticle association. It must be
emphasized, however, that this attractive interaction by itself does
\emph{not} reduce the problem to a standard case of electrostatic
stabilization resulting from an effective charge build-up on the colloids.
Instead, as is shown below, the resulting ``halo'' is dynamic in nature and
leads to effective interactions that are either attractive, repulsive, or
oscillatory, depending on nanoparticle concentration.

\begin{figure}
  \centering
  \includegraphics[width=\figurewidth]{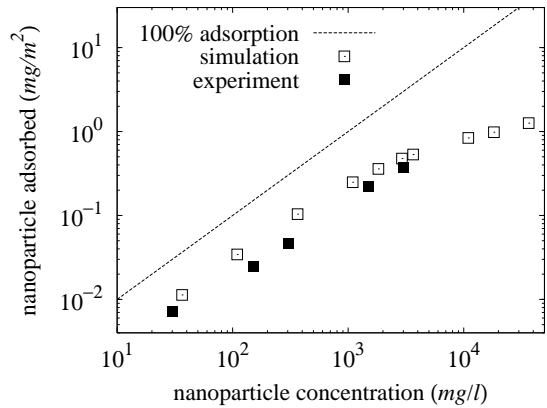}
  \caption{Nanoparticle adsorption per colloidal microsphere as a function of
  zirconia nanoparticle concentration, at fixed $\phi_{\rm micro} = 0.1$.
  The open squares represent the simulation results and the solid squares
  the experimental data~\protect\cite{tohver01}. $\phi_{\rm nano} = 10^{-3}$
  corresponds to a concentration of $3650$mg/l.}
  \label{fig:ads}
\end{figure}

The presence of a short-range attraction is indeed confirmed in
Fig.~\ref{fig:ads}, which provides a quantitative comparison of the Monte Carlo
simulations and the supernatant measurements.  The adsorption of nanoparticles
with diameter $\sigma_{\rm nano}=6{\rm nm}$ around a microsphere with diameter
$\sigma_{\rm micro} = 0.6\mu{\rm m}$ is determined as a function of
nanoparticle concentration, at fixed volume fraction $\phi_{\rm micro}=0.1$.
In order to account for colloidal many-body effects and to handle the extreme
size asymmetry ($\alpha = 100$), we use the generalized geometric cluster
algorithm~\cite{geomc} for a system containing 50 colloids and up to $5 \times
10^6$ nanoparticles.  In view of the small screening length and the presence of
short-range hydration forces, we employ the approximate expressions in Eqs.\ 
(\ref{eq:ele1}) and~(\ref{eq:ele2})~\cite{note-hydr}.  The amount of adsorbed
nanoparticles, which form a layer with an approximate thickness~$\sigma_{\rm
nano}$, is determined from the radial distribution function integrated from
contact to its first minimum.  While the Monte Carlo data to some extent
overestimate the adsorption, both data sets exhibit a linear dependence on
$\phi_{\rm nano}$ and a degree of adsorption that is far below 100\%. The
overestimation is compatible with the high zeta potential $\Psi_{\rm nano} =
70{\rm mV}$, somewhat above the regime of validity of the Debye--H\"uckel
approximation employed in the HHF equations~\cite{note-pb}. At volume fractions
above $10^{-3}$ the mutual repulsion between the charged nanoparticles causes
the adsorption to level off.

\begin{figure}
  \centering 
  \includegraphics[width=\figurewidth,angle=0]{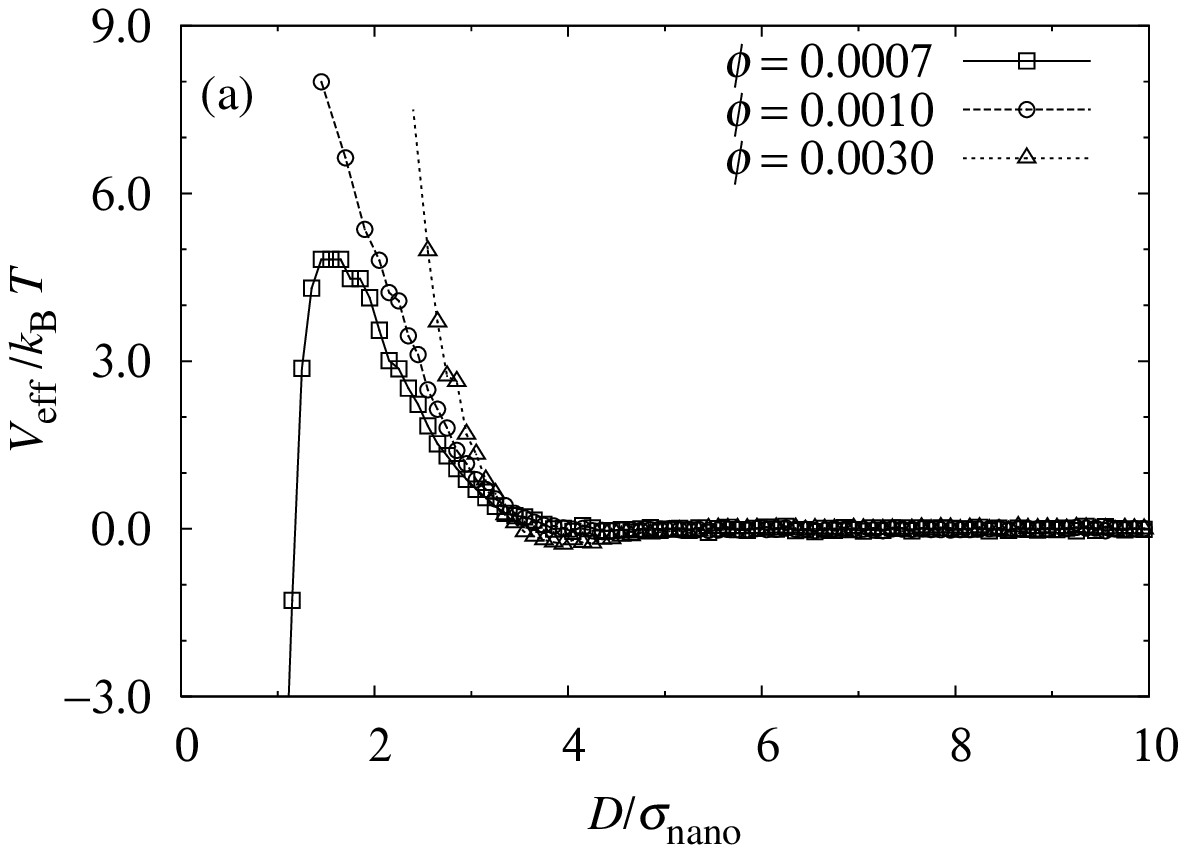}
  \includegraphics[width=\figurewidth,angle=0]{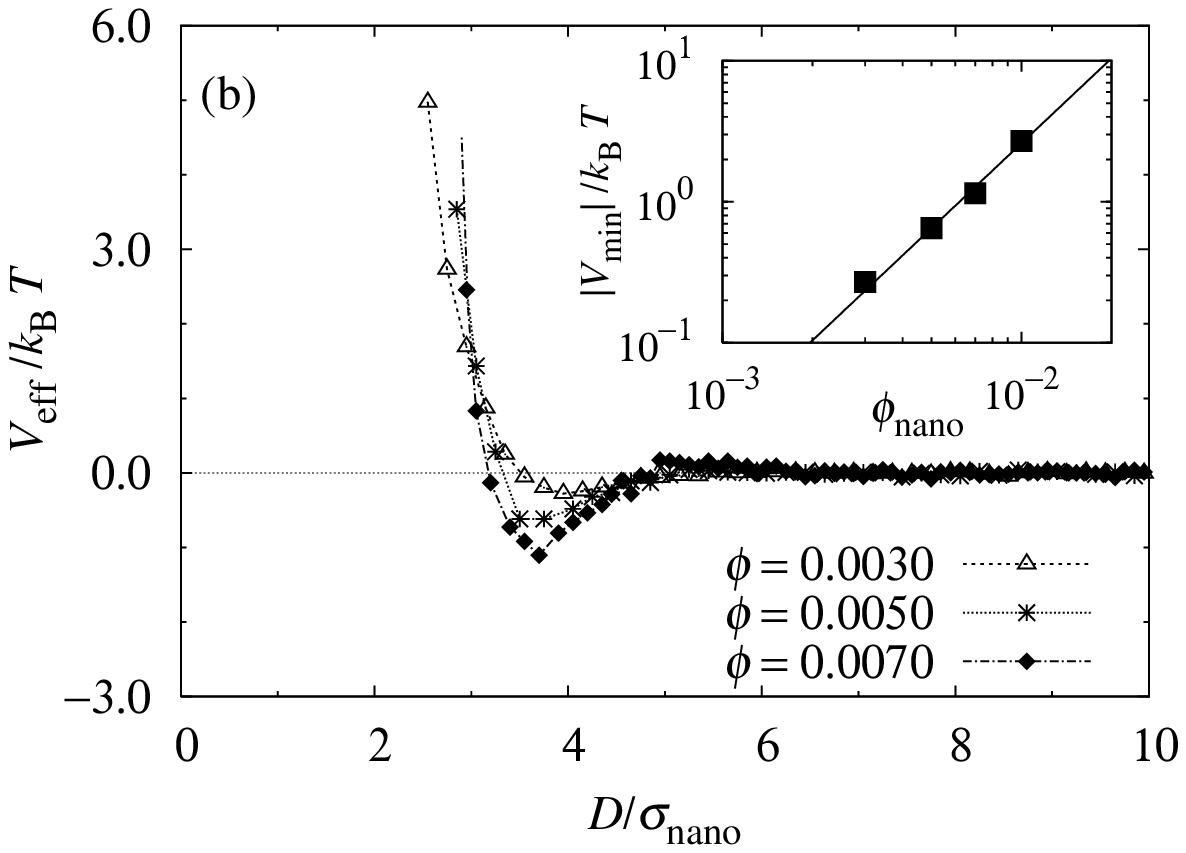}
  \caption{Effective potential $V_{\rm eff}/k_{\rm B}T$ between a pair of
  colloidal microspheres as a function of surface-to-surface separation~$D$.
  (a) At low nanoparticle volume fractions a bridging attraction is
  observed, which rapidly turns into a strong effective repulsion. (b) At
  higher volume fractions an attractive minimum develops, with a strength
  that grows \emph{quadratically} with $\phi_{\rm nano}$ (inset).}
  \label{fig:effective}
\end{figure}

Having established that the HHF equations model the experimental situation
relatively accurately, we proceed to determine the effective (induced)
colloidal interaction as a function of nanoparticle volume fraction. In the
infinite dilution limit the potential of mean force follows from the
colloidal pair correlation function, $V_{\rm eff} = -k_{\rm B}T \ln g(r)$.
The simulations employ a cubic box with $\phi_{\rm micro} =
0.002$~\cite{note-conc}.  As shown in Fig.~\ref{fig:effective}(a), at very
low $\phi_{\rm nano} = 7 \times 10^{-4}$ a strong attraction exists at a
separation~$\sigma_{\rm nano}$, accompanied by a repulsive barrier of
approximately $5k_{\rm B}T$. The potential well arises from nanoparticles
acting as ``linkers'' between two colloidal microspheres.  However, upon
increase of $\phi_{\rm nano}$ to $10^{-3}$ the bridging attraction rapidly
turns into a strong repulsion that extends over several nanoparticle
diameters. Since the direct van der Waals interaction has a strength of only
approximately $-1k_{\rm B}T$ at a surface separation $\sigma_{\rm
nano}$~\cite{tohver01b}, the total interaction exhibits a barrier.  This
repulsive barrier, which continues to increase with $\phi_{\rm nano}$, is
sufficiently strong to prevent gelation 
and thus leads to (kinetic) stabilization of the suspension.  The
corresponding threshold agrees remarkably well with the experimentally
observed gel--fluid transition near $\phi_{\rm nano} = 5 \times
10^{-4}$~\cite{tohver01}.

If the nanoparticle volume fraction is raised further, however, an
attractive well arises for $D \approx 4\sigma_{\rm nano}$
[Fig.~\ref{fig:effective}(b)]. As shown in the inset, the attraction
strength grows as $\phi_{\rm nano}^2$. At $\phi_{\rm nano} = 10^{-2}$ it
reaches nearly $-3 k_{\rm B}T$, which is estimated to be sufficient to
induce gelation.  Again, this packing fraction agrees quite well with the
experimentally observed reentrant gelation for $\phi_{\rm nano} \approx 5
\times 10^{-3}$.  The quadratic growth of the attraction strength with
$\phi_{\rm nano}$ indicates that this is not a regular depletion attraction,
which has a linear concentration dependence~\cite{asakura54}. Indeed,
examination of configurations shows only limited nanoparticle depletion
between colloidal pairs at $D \approx 4\sigma_{\rm nano}$.

\begin{figure}
  \centering
  \includegraphics*[width=.225\textwidth,angle=0]{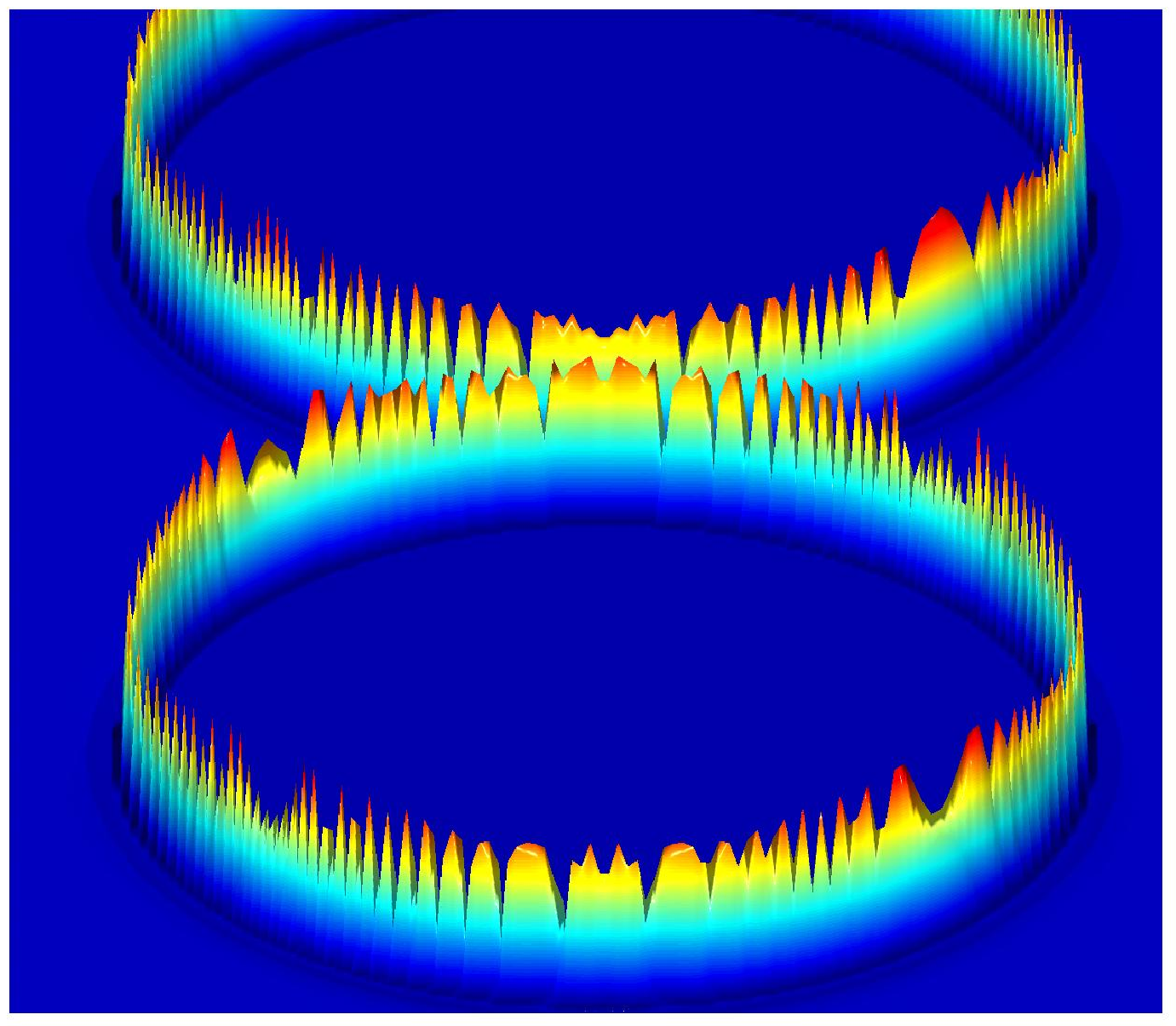}
  \includegraphics*[width=.225\textwidth,angle=0]{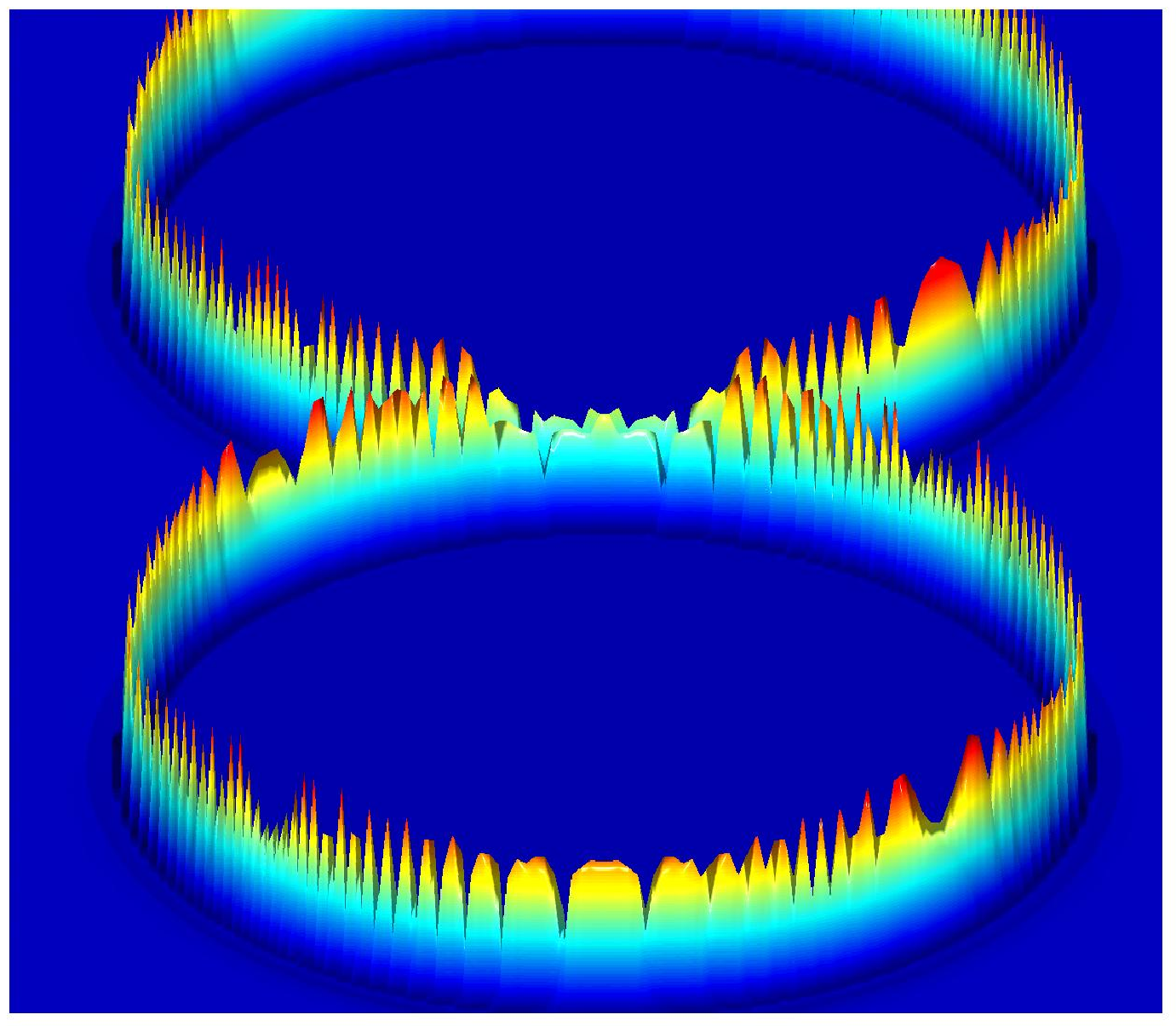}
  \caption{[color online] Nanoparticle density distribution around two
  colloidal particles at a surface-to-surface distance~$D$. (a) At larger
  separations ($D=5.0\sigma_{\rm nano}$) the halos around each colloid are
  uniform. (b) Upon closer approach ($D=1.8\sigma_{\rm nano}$), where there
  is a strong colloidal repulsion, a \emph{depletion} of nanoparticles in
  the gap region is observed.}
  \label{fig:distr}
\end{figure}

\begin{figure}
  \centering
  \includegraphics[width=\figurewidth,angle=0]{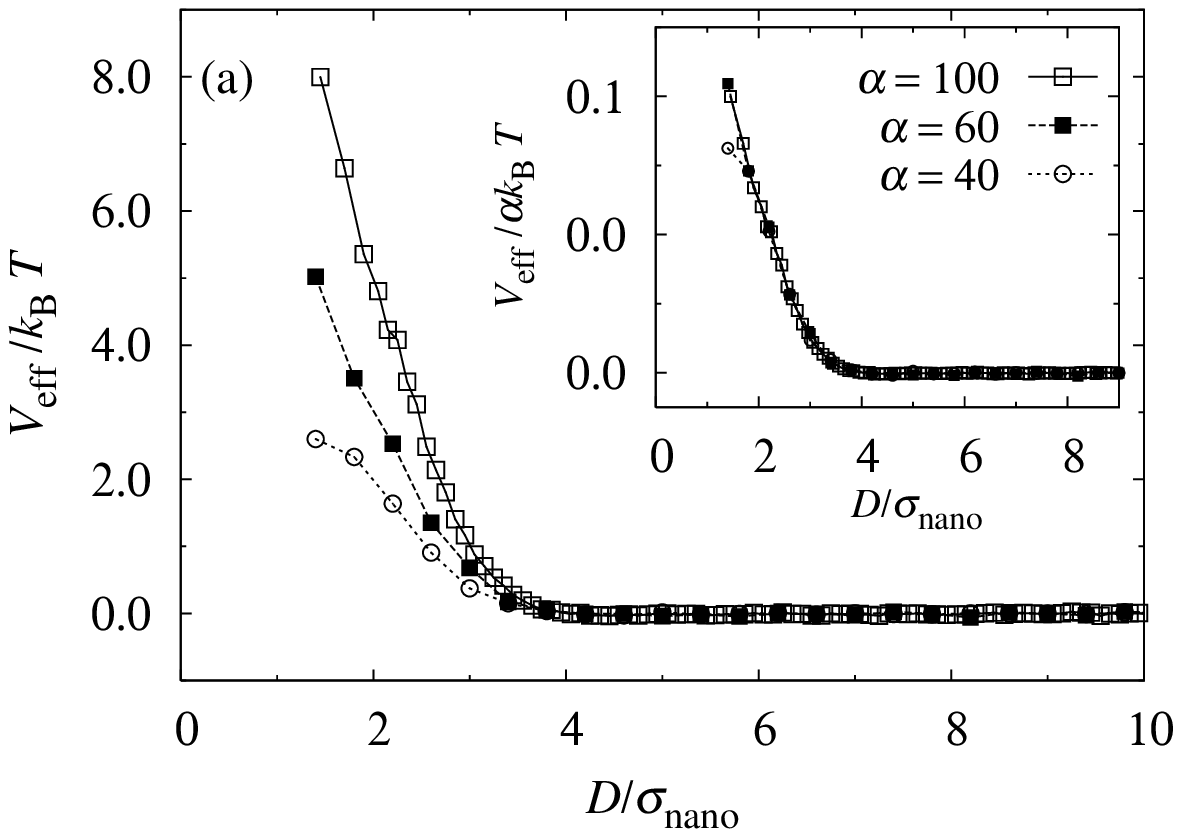}
  \includegraphics[width=\figurewidth,angle=0]{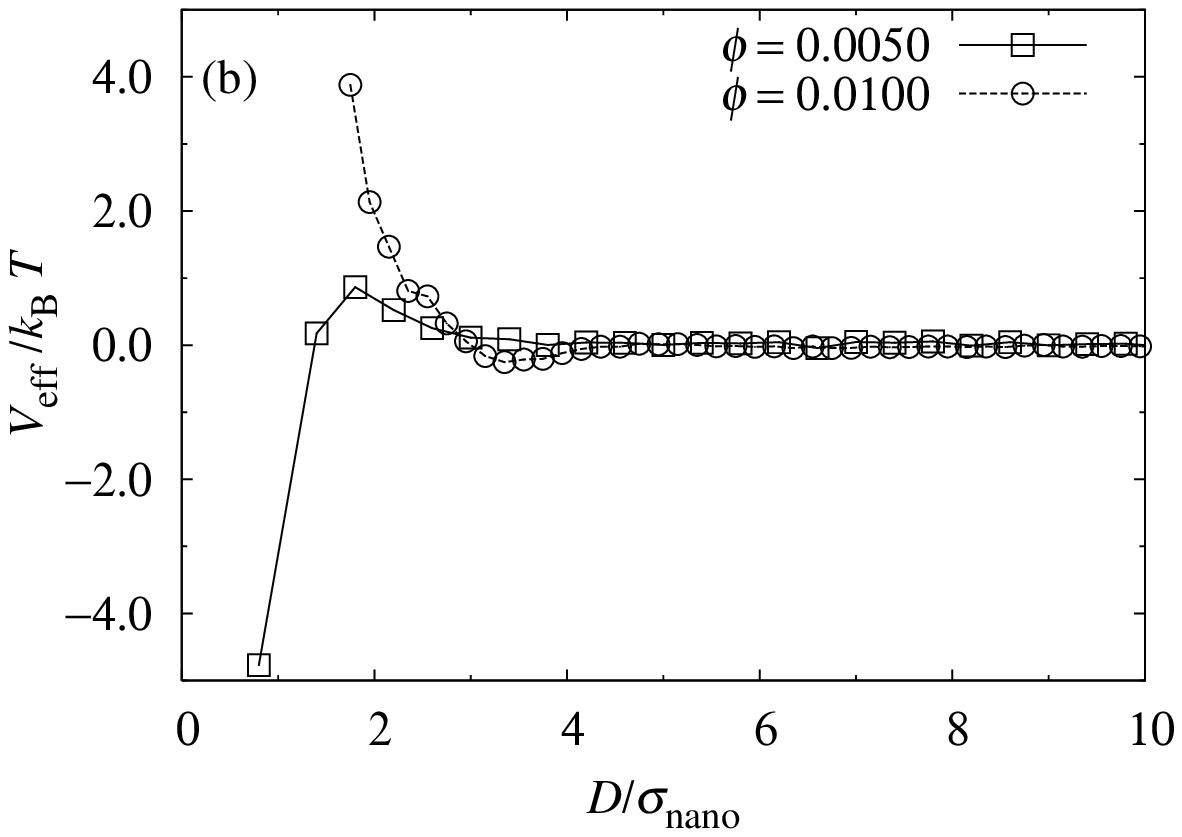}
  \caption{(a) Effect of nanoparticle--colloid size asymmetry~$\alpha$ on
  the effective colloidal interactions. At fixed nanoparticle size and
  volume fraction $\phi_{\rm nano}=0.001$, the repulsion increases with
  colloid size. As shown in the inset, the dependence is virtually perfectly
  linear.  (b) Effect of nanoparticle charge. All parameters are identical
  to those in Fig.~\protect\ref{fig:effective}, except for the nanoparticle
  surface potential, which has been lowered from $70{\rm mV}$ to $50{\rm
  mV}$. As a result, much larger nanoparticle volume fractions are required
  to achieve an appreciable colloidal repulsion.}
  \label{fig:parameters}
\end{figure}

In order to gain a microscopic understanding of the effective potentials we
study the distribution of nanoparticles around a pair of colloids, for various
separations of the colloidal particles. For clarity in the graphs, we employ a
smaller size ratio $\alpha=40$ (i.e., $\sigma_{\rm micro}=0.24\mu{\rm m}$) with
$\phi_{\rm nano}=0.003$, for which the effective interaction is strongly
repulsive [cf.\ Fig.~\ref{fig:parameters}(a) for $\phi_{\rm nano}=0.001$].  At
a surface separation $D \geq 5\sigma_{\rm nano}$ [Fig.~\ref{fig:distr}(a)] the
halos around both colloids are uniform, in agreement with a negligible
effective interaction. If the colloids are forced closer together, the mutual
repulsion between the nanoparticles leads to a redistribution of particles in
the halo. At a separation $D=1.8\sigma_{\rm nano}$ [Fig.~\ref{fig:distr}(b)]
most nanoparticles have been expelled from the gap region and a depletion zone
appears. While reminiscent of the depletion effect in hard-sphere
mixtures~\cite{asakura54}, the effective colloidal interaction in this case is
strongly \emph{repulsive} rather than attractive. This indicates that the
energy penalty resulting from nanoparticle rearrangements dominates over
depletion-like entropic effects.

The ``haloing'' stabilization mechanism clearly relies on rather generic
features and hence may become of considerable practical importance. Since
applications hinge on an understanding of the role of size and charge
asymmetry, we have investigated a variety of parameter combinations.
Figure~\ref{fig:parameters}(a) displays the repulsion resulting from a fixed
volume fraction $\phi_{\rm nano}=0.001$ for colloids of diameter
$0.60\mu{\rm m}$, $0.36\mu{\rm m}$, and $0.24\mu{\rm m}$ (size ratio $\alpha
= 100$, $60$, and $40$, respectively). Consistent with experimental
observation~\cite{tohver01}, the effective colloidal repulsion increases
with $\alpha$. As shown in the inset, the dependence on size asymmetry is
almost perfectly linear. The role of the nanoparticle \emph{charge} is
certainly complicated, as it affects both halo formation and the halo--halo
interaction.  Indeed, if the nanoparticle distribution around a pair of
colloids at surface separation~$D$ is denoted by $\rho(\mathbf{r'}; D)$,
where $\mathbf{r'}$ is measured with respect to the first colloid, then the
effective force on that colloid is
\begin{equation}
\mathbf{F}(D) = - \int \rho(\mathbf{r'};D)
\frac{\partial}{\partial \mathbf{r'}} V_{\rm micro-nano}(r')d\mathbf{r'} \;.
\label{eq:eff_force}
\end{equation}
The nanoparticle charge will affect both $\rho(\mathbf{r'}; D)$ and $V_{\rm
micro-nano}$. In the absence of experimental data, we have performed
simulations for the $\alpha=100$ case with a lower nanoparticle surface
potential, $\Psi_{\rm nano} = 50{\rm mV}$ [Fig.~\ref{fig:parameters}(b)].
The effective interactions are found to be similar to those for higher
surface potentials, with a comparable dependence on $\phi_{\rm nano}$, but
almost ten times larger volume fractions are needed to achieve the same
effective colloidal repulsion. This emphasizes the important role of the
nanoparticle charge. Finally, we have performed control simulations in which
the colloid--nanoparticle attraction is omitted. Neither the observed
adsorption (Fig.~\ref{fig:ads}) nor the effective repulsions
(Fig.~\ref{fig:effective}) can be reproduced unless the nanoparticle
concentrations are \emph{much} higher than in the experiment or the
nanoparticle repulsion has a much longer range, which in turn would be
incompatible with the actual screening length $\kappa^{-1} \approx
\sigma_{\rm nano}/3$~\cite{tohver01b}.

In summary, the mechanism for colloidal stabilization through nanoparticle
``haloing''~\cite{tohver01} has been clarified by means of direct simulation.
The geometric cluster algorithm, which provides an acceleration of many orders
of magnitude~\cite{geomc}, has been indispensable for this: At a nanoparticle
volume fraction $\phi_{\rm nano}=0.007$ the calculation involves $7 \times
10^6$ nanoparticles, requiring a simulation of $\mathcal{O}(10^4)$ hours.  We
have shown that an induced colloid--nanoparticle attraction contributes to the
effective repulsion, but that it by no means reduces the situation to simple
steric or Coulombic stabilization. Haloing stabilization occurs over a window
of nanoparticle concentrations, bracketed by two dissimilar gel phases, neither
of which is a regular depletion gel.  Our calculations indicate that the
effective colloidal interactions also sensitively depend on nanoparticle charge
and colloid--nanoparticle size asymmetry, so that the haloing mechanism indeed
opens up a variety of opportunities to tailor materials properties.

\begin{acknowledgments}
  We gratefully acknowledge stimulating discussions with Jennifer Lewis and
  Ken Schweizer.  This material is based upon work supported by the U.S.
  Department of Energy, Division of Materials Sciences under Grant No.\
  DEFG02-91ER45439, through the Frederick Seitz Materials Research
  Laboratory at the University of Illinois at Urbana-Champaign and by the
  National Science Foundation under Grants No.\ DMR-0346914 and CTS-0120978.
\end{acknowledgments}

Note: Upon completion of this work we became aware of a study by
Karanikas and Louis~\cite{karanikas04}, who use different techniques, but
arrive at similar conclusions.


\begin{thebibliography}{17}
\bibitem[{not({\natexlab{c}})}]{note-conc}
\bibinfo{note}{We have verified that the results remain unchanged upon further
  decrease of the colloid volume fraction.}
\end{thebibliography}
\end{document}